\documentclass[journal]{IEEEtran}

\usepackage{xcolor,soul,framed} 
\usepackage{amssymb,stmaryrd}   
\colorlet{shadecolor}{yellow}
\usepackage[pdftex]{graphicx}
\graphicspath{{../pdf/}{../jpeg/}}
\DeclareGraphicsExtensions{.pdf,.jpeg,.png}
\usepackage{bm}
\usepackage[cmex10]{amsmath}
\usepackage{array}
\usepackage{mdwmath}
\usepackage{mdwtab}
\usepackage{eqparbox}
\usepackage{ur  l}
\usepackage{hyperref}
\usepackage{cite}
\usepackage{amssymb,amsfonts}
\usepackage{algorithmic}
\usepackage{graphicx}
\usepackage{textcomp}
\usepackage{xcolor}
\usepackage{multirow} 
 \usepackage{booktabs}
 \usepackage{graphicx} 
 \usepackage{graphicx, caption, sidecap}
\def\BibTeX{{\rm B\kern-.05em{\sc i\kern-.025em b}\kern-.08em
    T\kern-.1667em\lower.7ex\hbox{E}\kern-.125emX}}
\usepackage{array, makecell} %
\usepackage{stmaryrd}
\usepackage{tikz}    
\usepackage{enumerate}
\usepackage{tikz}
\usepackage{amsmath}
\usetikzlibrary{positioning, calc, shapes.geometric}

\newtheorem{prop}{Proposition}

\newenvironment{proof}{\textit{Proof}: }

\newcommand{\fig}[4]{ \begin{figure}[#4]
  \centering
   \includegraphics[width=#3\textwidth]{Figures/#1}
   \caption{#2}\label{fig:#1}
  \end{figure}
}

\begin{document}

\title{Practical Short-Length Coding Schemes for  Binary Distributed Hypothesis Testing}

\author{
 Ismaila Salihou Adamou, Elsa Dupraz, \textit{Senior Member, IEEE}, Reza Asvadi, \textit{Senior Member, IEEE}, and Tadashi Matsumoto, \textit{Life Fellow, IEEE}
\thanks{Ismaila Salihou Adamou and Elsa Dupraz are with the Department of Mathematical and Electrical Engineering, IMT Atlantique, 29238 Brest, France}
\thanks{Reza Asvadi is with the Department of Telecommunications, Faculty of
Electrical Engineering, Shahid Beheshti University, Tehran 1983969411, Iran}
\thanks{Tadashi Matsumoto is with the Department of Mathematical and Electrical Engineering, IMT Atlantique, 29238 Brest, France, and also with the Centre for Wireless Communications,
University of Oulu, 90014 Oulu, Finland. He is also Professor Emeritus
of Japan Advanced Institute of Science and Technology, Japan}
\thanks{ This work has received French government support granted to the Cominlabs excellence laboratory and managed by the National Research Agency in the ``Investing for the Future'' program under reference ANR-10-LABX-07-01. }
\thanks{Part of the content of this paper was published in: Elsa Dupraz, Ismaila Salihou Adamou, Reza Asvadi, Tadashi Matsumoto, Practical Short-Length Coding Schemes for Binary Distributed Hypothesis Testing, \emph{International Symposium on Information Theory (ISIT)}, 2024. We extend the analysis to encompass the symmetric two encoders hypothesis testing setup. }
}

\maketitle 

\begin{abstract}
This paper addresses the design of practical short-length coding schemes for Distributed Hypothesis Testing (DHT). While most prior work on DHT has focused on information-theoretic analyses—deriving bounds on Type-II error exponents via achievability schemes based on quantization and quantize-binning—the practical implementation of DHT coding schemes has remained largely unexplored. 
Moreover, existing practical coding solutions for quantization and quantize-binning approaches were developed for source reconstruction tasks considering very long code length, and they are not directly applicable to DHT. 
In this context, this paper introduces efficient short-length implementations of quantization and quantize-binning schemes for DHT, constructed from short binary linear block codes. Numerical results show the efficiency of the proposed coding schemes compared to uncoded cases and to existing schemes initially developed for data reconstruction.  
In addition to practical code design, the paper derives exact analytical expressions for the Type-I and Type-II error probabilities associated with each proposed scheme. The provided analytical expressions are shown to predict accurately the practical performance measured from Monte-Carlo simulations of the proposed schemes. These theoretical results are novel and offer a useful framework for optimizing and comparing practical DHT schemes across a wide range of source and code parameters.

\end{abstract}


\section{Introduction}
In the era of 5th generation (5G) wireless communications system and beyond, the paradigm of communication systems is shifting to address emerging challenges and requirements. Historically, these systems were designed to ensure reliable transmission,  mostly focusing on minimizing error probability or distortion between original and reconstructed data~\cite{wyner1976rate, gray1998quantization}.  However, modern communication systems are now increasingly dedicated to specific tasks that require optimized design. Especially in the emerging field of goal-oriented communications \cite{strinati20216g, stavrou2022rate,zou2022goal}, the objective is no longer to reconstruct the data but rather to apply specific tasks directly upon the received data. While examples of such tasks include regression,  classification, or semantic analysis,  
this paper focuses on the important case of decision-making. Examples of applications include sensors embedded in the human body for health disease detection~\cite{linden2018embedded}, underwater activity monitoring~\cite{lior2021detection}, or traffic jam detection from route planning of autonomous vehicles~\cite{sanchez2024distributed}. 

In Information Theory, the problem of decision-making over coded data is formalized as distributed hypothesis testing (DHT), first introduced in~\cite{ahlswede1986hypothesis}. DHT considers a scenario with two separate terminals: one observing a source $X$, and the other observing a source $Y$. These sources are jointly distributed according to the joint distribution $P_{XY}$, which depends on one of the two hypotheses, $\mathcal{H}_0$ or $\mathcal{H}_1$~\cite{ahlswede1986hypothesis, Han1987, amari1998statistical}. 
The terminals transmit encoded messages under given rate constraints, $R_1$ for $X$ and $R_2$ for $Y$. The receiver then makes a decision between $\mathcal{H}_0$ and $\mathcal{H}_1$, by  applying an hypothesis test over the received coded data. 
In this work, we focus on two distinct setups. The first setup, referred to as the asymmetric setup, assumes that the receiver has access to a coded version of $X$, while $Y$ is losslessly observed as side information at the receiver. This setup is similar to the Wyner-Ziv setup for lossy source coding~\cite{wyner1976rate}. The second setup, known as the symmetric setup, assumes that both $X$ and $Y$ are encoded. 

In both setups, the hypothesis testing performance is characterized by two types of error probabilities: Type-I error probability, denoted by $\alpha_n$, and Type-II error probability, denoted by $\beta_n$. A Type-I error occurs when $\mathcal{H}_1$ is chosen while the true hypothesis is $\mathcal{H}_0$, whereas a Type-II error arises when $\mathcal{H}_0$ is selected while $\mathcal{H}_1$ is true. The central question in the framework of DHT is: how can one design coding schemes so as to satisfy the rate constraints while ensuring optimal decision-making at the receiver? Here, optimality is defined as minimizing the Type-II error probability, $\beta_n$, under a given constraint on the Type-I error probability $\alpha_n$. Addressing this question requires tackling both the Information-Theoretic limits of the problem and the practical design of coding schemes.

\subsection{Prior Works in Information Theory}
In the Information-Theoretic analysis of DHT, the primary objective is to characterize the achievable error exponent of the Type-II error probability, while the Type-I error probability is kept below
a prescribed threshold~\cite{ahlswede1986hypothesis,Han1987}. The asymmetric setup has been extensively studied, with several achievable coding schemes proposed to refine lower bounds on the error exponent. 
Ahlswede and Csiszár first introduced the quantization scheme\cite{ahlswede1986hypothesis}, which is optimal for a special case known as testing against independence. 
Han enhanced the quantization scheme by incorporating a joint typicality check between the source and its quantized version\cite{Han1987}. However, these approaches do not fully exploit the correlation between sources $X$ and $Y$. 

To address this limitation, Shimokawa et al. proposed the Shimokawa-Han-Amari (SHA) scheme, also known as the quantize-binning scheme, which performs random binning after the quantization~\cite{shimokawa1994error}. This approach originates from the Wyner-Ziv coding scheme~\cite{wyner1976rate} and achieves tighter lower bounds on the error exponent. Further refinements to the SHA scheme addressed the trade-off between binning errors and hypothesis testing errors, as investigated for i.i.d. sources in~\cite{Katz2015} and for non-i.i.d. sources in~\cite{adamou2023information}. Most recently, Kochman and Wang enhanced the SHA scheme by refining the entropy check introduced by Shimokawa et al.~\cite{kochman2023improved}.
Finally, the SHA scheme has been generalized to more complex communication scenarios, including discrete memoryless channels\cite{sreekumar2019distributed}, multiple-access channels\cite{salehkalaibar2018distributed}, and two-hop relay networks~\cite{salehkalaibar2019hypothesis}.


The symmetric setup has received much 
less attention, except in the specific case of zero-rate compression where one or both coding rates asymptotically approach
zero~\cite{Han1987, shalaby1992multiterminal, han2006exponential, amari1989statistical, watanabe2017neyman}. The zero-rate case has limited relevance for conventional lossless or lossy compression, but it has important applications in statistical hypothesis testing~\cite{Han1987}. In this case, research has focused on designing optimal testing schemes~\cite{watanabe2017neyman, Haim2017} and characterizing achievable error exponents~\cite{Han1987, shalaby1992multiterminal, han2006exponential, amari1989statistical, watanabe2017neyman, Haim2017}.

\subsection{Prior works on practical coding schemes}
While the Information-Theoretic performance of DHT has been extensively studied, design of practical coding schemes for this setup has received considerably less attention. In fact, Information-Theoretic schemes are not directly implementable and largely rely on unpractical assumptions, such as infinite block lengths, which are incompatible with the finite and typically short sequences encountered in practical decision-making scenarios. This gap in the literature serves as the primary motivation of this paper, with a focus on binary sources.

For data reconstruction, several practical coding techniques have been proposed in the literature. They include binary quantizers~\cite{fridrich2007binary}, binning schemes~\cite{xiong2004distributed, liveris2002compression, savard2013improved,ye2019optimized}, and quantize-binning schemes~\cite{wainwright2009low}, all constructed using linear block codes. For instance, in~\cite{liveris2002compression, savard2013improved}, binning schemes based on Low-Density Parity-Check (LDPC) codes are proposed, achieving lossless compression near the Slepian-Wolf limit for correlated binary sources. Similarly, binary quantization schemes have been developed using LDPC codes~\cite{matsunaga2003coding}, or Low-Density Generator Matrix (LDGM) codes of which decoding is conducted by Bias Propagation (BiP) algorithms~\cite{wainwright2010lossy, fridrich2007binary,castanheira2010lossy}. Furthermore, it was shown in~\cite{wainwright2009low, nangir2018binary} than compound LDGM/LDPC constructions for quantize-binning schemes can achieve the Wyner-Ziv rate-distortion function for binary sources. 

While these coding schemes are effective for source reconstruction, they are less suitable for DHT due to their use of belief propagation decoding algorithm which are designed for very long sequences, often exceeding $10^3$ to $10^4$ bits. In contrast, DHT involves short-length sequences, where only a few dozen bits may suffice for accurate decision-making. 
This raises an important question on how to design  efficient quantizers and quantize-binning schemes with short length codes. 
Another key issue is the design of effective hypothesis tests over coded data, given that the methods proposed in Information-Theoretic proofs of DHT are impractical. In this paper, we address these issues and propose efficient short-length coding schemes for DHT.


\subsection{Contributions}
In this paper, we introduce a first practical short-length coding scheme for DHT in the asymmetric setup, where side information $Y$ is fully available at the decoder. We then extend this scheme to the symmetric setup, where both the source $X$ and side information $Y$ are compressed by independent  encoders. We propose practical quantization and quantize-binning schemes  for DHT in both asymmetric and symmetric setups.

Based on the principles of practical Wyner-Ziv coding, our schemes employ linear block codes designed specifically for short block lengths ($n<100$ bits). 
We first describe the construction of the coding scheme and provide the hypothesis test expression. We then derive exact analytical expressions for the Type-I and Type-II error probabilities for the code given. These analytical tools are novel, and enable the optimization and comparison of the proposed schemes across a broad range of source and code parameters. 

The major contributions of this paper are summarized as follows.
\begin{itemize}
\item We discuss and compare two uncoded schemes for DHT. In the first scheme, called the separate scheme, each encoder independently makes a local decision based on its observation and transmits a single bit to the receiver. 
In the second scheme, called the truncation scheme, the encoders transmit only the first $l<n$ bits of their source sequences to the receiver. While these schemes are not novel, they provide reference points for evaluating the proposed quantizer and quantize-binning schemes. They also allow us to compare decide-and-compress versus compress-and-decide strategies, in line with previous Information-Theoretic works that investigated estimate-and-compress versus compress-and-estimate setups~\cite{wolf1970transmission,liu2021rate}. 

\item For both symmetric and asymmetric setups, we introduce quantizer-alone and quantize-binning schemes for DHT, constructed with short-length linear block codes ($n<100$). Simulation results show that the proposed constructions outperform solutions initially proposed for long block-length in~\cite{wainwright2010lossy, fridrich2007binary,castanheira2010lossy}. The simulation results also demonstrate the superiority of the quantize-binning scheme over the quantizer-alone and uncoded truncation schemes, particularly when code parameters are optimized.
\item We derive exact analytical expressions for the Type-I and Type-II error probabilities of the quantizer-alone and quantize-binning schemes in the asymmetric setup.  Numerical results validate the accuracy of the analytical error probability expressions by comparison with Monte-Carlo simulations. 
\end{itemize}

\subsection{Outline}
The remainder of this paper is organized as follows. Section~\ref{sec:system_model} presents the DHT setup.  
Section~\ref{sec:uncoded} introduces the uncoded schemes. 
Section \ref{sec:quant} presents the quantization scheme and provides the analytical expressions for Type-I and Type-II
error probabilities. Section~\ref{sec:quant_bin} describes the quantize-binning scheme and provides the analytical expressions for Type-I and Type-II error probabilities. 
Section~\ref{sec:numerical_results} presents the numerical results. 


    \section{Distributed Hypothesis Testing}\label{sec:system_model}

    This section introduces the considered DHT setup and presents existing Information-Theoretic coding schemes for this problem. 
    
    \subsection{Notation}
Let $\llbracket 1,M \rrbracket$ denote the set of integers from $1$ to $M$. Random variables are represented by uppercase letters, \emph{e.g.}, $X$, while their realizations are in lowercase, \emph{e.g.}, $x$. 
Boldface letters, \emph{e.g.}, $\mathbf{X}^n$ indicate vectors of length $n$. 
The Hamming weight of a vector $\mathbf{x}^n$ is denoted as $w(\mathbf{x}^n)$, and the Hamming distance between two vectors $\mathbf{x}^n$ and $\mathbf{y}^n$ is $d(\mathbf{x}^n, \mathbf{y}^n)$. The binomial coefficient of two integers $n,  k$,    with $k \leq n$, is expressed as $\binom{n}{k}$.
    
    \subsection{Hypothesis Testing with binary sources}\label{subsec:binary_sources}
We consider two source vectors $\mathbf{X}^n$ and $\mathbf{Y}^n$ of length $n$. Like in the conventional DHT setup~\cite{Han1987}, we assume that the components of $\mathbf{X}^n$ and $\mathbf{Y}^n$ are i.i.d., drawn according to random variables $X$ and $Y$, respectively. The pair $(X,Y)$ follows one of two possible joint distributions:
    \begin{align}
        \mathcal{H}_0  &: \left(X,Y\right) \sim \mathbb{P}_{0}, \notag \\
         \mathcal{H}_1  &: \left(X,Y\right) \sim \mathbb{P}_{1}.
        \label{eq:general_dht}
    \end{align}
In what follows, with a slight abuse of notation, we always denote $\mathbb{P}_0$ (resp. $\mathbb{P}_1$) the random variable or vector probability distribution under $\mathcal{H}_0$ (resp. $\mathcal{H}_1$).  For instance,   $\mathbb{P}_0(\mathbf{x}^n)$ is the probability of vector $\mathbf{x}^n$ under $\mathcal{H}_0$. 
    We consider a general case where the marginal distributions of $X$  under $\mathcal{H}_0$ and $\mathcal{H}_1$ are not necessarily identical. Therefore, the marginal distributions of $Y$  under $\mathcal{H}_0$ and $\mathcal{H}_1$,  are not necessarily  identical neither. 
    
    In this work, we focus on binary sources, where $X$ and $Y$ take values in the alphabet $\{0,1\}$ and follow the model $Y = X \oplus Z$, with $Z$ being a binary random variable independent of $X$. Let $p = \mathbb{P}(X=1)$ and $c = \mathbb{P}(Z=1)$, where $0 < c \leq 1/2$. The hypotheses provided in~\eqref{eq:general_dht} can then be expressed as:
    \begin{align}
    \mathcal{H}_0: & (p=p_0, c=c_0) \notag\\ \mathcal{H}_1: & (p=p_1,  c=c_1).
    \label{binary_dht}
    \end{align}
For convenience, we assume that $0 < p_0 \leq 0.5$, $p_0 \leq p_1$, and $c_0 \leq c_1$. This model has been studied from an Information-Theoretic perspective in~\cite{Katz2017, Haim2017}. Notably, when $p_0 = p_1$ and $c_1 = 1/2$, the problem reduces to testing against independence~\cite{ahlswede1986hypothesis}.
    
    \subsection{Coding schemes}

    Figure~\ref{fig:system_two_enc_ht} illustrates the DHT setup. Encoder $1$ and Encoder 2 send coded representations of $\mathbf{X}^n$ and $\mathbf{Y}^n$ at rates $R_1$ and $R_2$, respectively. The decoder uses the received coded information to make a decision between $\mathcal{H}_0$ and $\mathcal{H}_1$. In this paper, two setups are considered:
    \begin{itemize}
        \item \textit{Asymmetric setup}: $\mathbf{X}^n$ is encoded at rate $R_1$, while $\mathbf{Y}^n$ is fully available at the decoder.
        \item \textit{Symmetric setup}: both $\mathbf{X}^n$ and $\mathbf{Y}^n$ are encoded at rates $R_1$ and $R_2$, respectively.
    \end{itemize}

   \fig{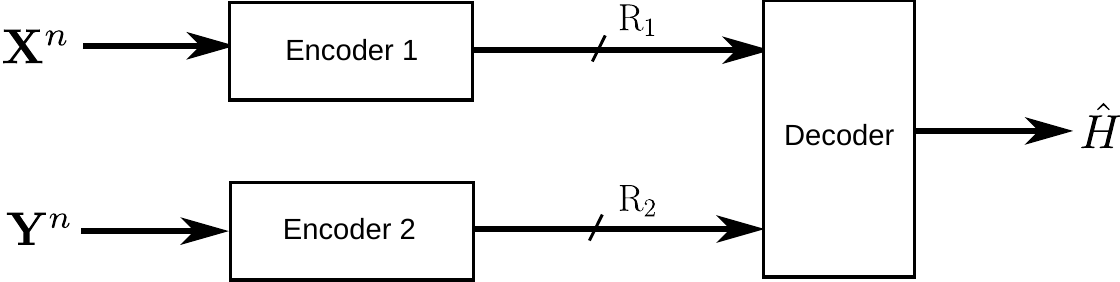}{Distributed hypothesis testing scheme}{0.48}{t}
    
    For a given block length $n$, the encoding functions are defined as:
    \begin{align}
     f_{1}^{(n)}&:\{0,1\}^n \rightarrow \llbracket 1,2^{nR_1} \rrbracket , \\
    f_{2}^{(n)}&: \{0,1\}^n \rightarrow \llbracket 1,2^{nR_2} \rrbracket,
    \end{align}
    and the decision function as:
    \begin{equation}
        g^{(n)}:\llbracket 1,2^{nR_1} \rrbracket\times\llbracket 1,2^{nR_2} \rrbracket\rightarrow \{0,1\} .
    \end{equation}
    
   For given encoding and decision functions $(f_{1}^{(n)}, f_{2}^{(n)}, g^{(n)})$, we define Type-I error probability $\alpha_n$ and Type-II error probability $\beta_n$ as~\cite{Haim2017} 
    \begin{align}
     \alpha_n & = \mathbb{P}_0\left(g^{(n)}(  f_{1}^{(n)}(\mathbf{X}^n), f_{2}^{(n)}(\mathbf{Y}^n)  ) = 1\right) , \\
     \beta_n & = \mathbb{P}_1\left(g^{(n)}(  f_{1}^{(n)}(\mathbf{X}^n), f_{2}^{(n)}(\mathbf{Y}^n)  ) = 0 \right) . 
    \end{align}
    For a given value $\epsilon \in (0,1)$ such that $\alpha_n < \epsilon$, Type-II error exponent $\theta$ is defined as~\cite{Haim2017} 
    \begin{equation}\label{eq:def_ee}
     \lim_{n \rightarrow \infty} \sup \frac{1}{n} \log \frac{1}{\beta_n} \geq \theta.
    \end{equation}
    
    \subsection{Short-length nature of DHT}
Existing lower bounds on the error exponent $\theta$ are obtained by considering $n \to \infty$. However, they provide scaling laws for $\beta_n$, approximating $\beta_n \approx e^{-n\theta}$. Here, we use this approximation to investigate the block length $n$ needed to achieve sufficiently low Type-II error probability. 

Among existing lower bounds on the error-exponent for the asymmetric setup, the one provided in~\cite{Katz2017} is based on a quantize-binning achievable coding scheme and is simpler to evaluate than the one of the SHA scheme in~\cite{shimokawa1994error}. 
    Specifying this bound for the DHT problem with binary sources defined in~\eqref{binary_dht} leads to 
    \begin{align}\label{eq:theta_bin_kp}
     \theta  \geq & \sup_{\delta \in [0,1]}  \min \bigg\{ R_1-[H_{2}(p_0*\delta)-H_{2}(\delta)],  \\
     & \left. (p_0*\delta)\log\frac{p_0*\delta}{p_1*\delta}+(1-(p_0*\delta))\log\frac{1-(p_0*\delta)}{1-(p_1*\delta)} \right\}. \notag
    \end{align}
    Here, $H_2$ is the binary entropy function, and $*$ is the binary convolution operator defined as $ x * y = (1-x)y + (1-y)x $, with $0 \leq x,y \leq 1$. In addition, $\delta$ is a parameter of the Information-Theoretic coding scheme. 

For instance, with $p_0 = 0.05$, $p_1 = 0.5$, $\delta = 0.1$, and $R_1 = 0.4$, $\beta_n$ drops from $10^{-6}$ with $n=50$ to $10^{-12}$ with $n=100$. This highlights the importance of focusing on small $n$ values ($n < 100$), motivating our study of short-length coding schemes.

    \subsection{Information-Theoretic coding scheme}
    We now review the SHA quantize-binning achievable  scheme which have been proposed in the literature of DHT for the asymmetric setup~\cite{shimokawa1994error}. This will allow us to identify the main steps of a practical coding for this problem. For brievety, we do not provide here the corresponding lower bound on the error exponent. Indeed, the bound has a quite complex expression and is not essential to convey the main message of this part. 
    The quantize-binning scheme of~\cite{shimokawa1994error},\cite{rahman2012optimality} operates as follows.
    \begin{itemize}
        \item \textbf{Codebook generation}: $2^{nr_1}$ sequences $\mathbf{u}^n$ are generated at random according to a pre-defined distribution $P_{U|X}$. Those sequences $\mathbf{u}^n$  are then distributed randomly and uniformly into $2^{nR_1}$ bins, with $R_1 < r_1$.
        \item \textbf{Encoding}: to encode $\mathbf{x}^n$, the encoder first selects a codeword $\mathbf{\hat{u}}^n$ that is jointly typical with $\mathbf{x}^n$~\cite{shimokawa1994error}. 
    The codeword $\mathbf{\hat{u}}^n$ can be interpreted as a quantized version of $\mathbf{x}^n$. The encoder then sends to the decoder the index $i \in \llbracket 1, 2^{nR_1} \rrbracket$ of the bin to which the sequence $\mathbf{\hat{u}}^n$ belongs.
    \item \textbf{Hypothesis test}:  the receiver uses an empirical entropy check~\cite{shimokawa1994error} to extract the sequence $\mathbf{\hat{u}}^n$ in the bin $i$ that has the lowest empirical entropy $H_e(\mathbf{u}^n|\mathbf{y}^n)$ defined as
    \begin{equation}\label{eq:empirical_entropy}
    H_e(\mathbf{u}^n|\mathbf{y}^n) = -\frac{1}{n} \sum_{k=1}^n \log P_{U|Y}(u_k |y_k), 
    \end{equation}
    where $P_{U|Y}$ is the conditional probability distribution of $U$ given $Y$ under $\mathcal{H}_0$. 
    The receiver then declares $\mathcal{H}_0$ if the extracted sequence $\mathbf{\hat{u}}^n$ and the side information $\mathbf{y}^n$ are jointly typical.  
    \end{itemize}
   The Information-Theoretic coding scheme described above relies on quantization and binning steps, which we aim to implement in a practical form. 
   In this work, we will rely on linear block codes for both steps. 
   
   Next, the previous criteria of joint typicality  and empirical entropy in~\eqref{eq:empirical_entropy} allow some error probability terms to vanish asymptotically in the Information-Theoretic proof. In this work, we consider the finite-length regime. Consequently, we adopt the Maximum Likelihood (ML) estimator and the Neyman-Pearson (NP) test, which are standard techniques in signal processing and are known to be optimal under specific conditions, as detailed later in the paper.

\section{Uncoded schemes} \label{sec:uncoded}
Before introducing the proposed practical quantize and quantize-binning schemes, in this section we describe two schemes that do not require coding. 
These schemes will serve as baselines when evaluating the performance of our proposed coding schemes.  

\subsection{Separate scheme (local decisions)}\label{sec:alpha_beta_sep}
When the marginal distributions of $X$ and $Y$ depend on the hypothesis $\mathcal{H}_0$ or $\mathcal{H}_1$ (that is $p_0  \neq p_1$), hypothesis tests can be constructed directly at the encoders based only on local observations $\mathbf{x}^n$ for Encoder $1$ and $\mathbf{y}^n$ for Encoder $2$. 
Each encoder will then send a single bit to inform the decoder of its decision. 
This setup offers the advantage of achieving very low communication rates by transmitting just one bit of information to the decoder. This constitutes a special case of the estimate-and-compress setup initially introduced in~\cite{wolf1970transmission} in the context of parameter estimation from compressed data. 
 
 In the separate scheme, the encoding functions $f_{1}^{(n)} : \{0,1\}^n \to \{0,1\}$ and $f_{2}^{(n)}$:$ \{0,1\}^n \to \{0,1\}$ are constructed based on a NP test~\cite{lehmann2005testing} on the observations $\mathbf{x}^n$, and $\mathbf{y}^n$, respectively.
The coding rates are given by $R_1 = R_2 = 1/n$. 
%
Under the constraints \( \alpha_n^{(1)} < \epsilon \) and \( \alpha_n^{(2)} < \epsilon \) on the Type-I error probabilities for Encoder~1 and Encoder~2, respectively, the Neyman-Pearson (NP) lemma~\cite{lehmann2005testing} provides optimal decision rules. Specifically, the following tests:
\begin{align}
\mathbb{P}_{1}(\mathbf{x}^n) &< \mu_1  \mathbb{P}_{0}(\mathbf{x}^n), 
\label{eq:np_enc1}
\\
\mathbb{P}_{1}(\mathbf{y}^n) &< \mu_2  \mathbb{P}_{0}(\mathbf{y}^n), 
\label{eq:np_enc2}
\end{align}
where in each case $\mathcal{H}_0$ is decided if the inequality is satisfied, and $\mathcal{H}_1$ is decided otherwise, 
 minimize the Type-II error probabilities \( \beta_n^{(1)} \) and \( \beta_n^{(2)} \), respectively. Here, the thresholds \( \mu_1 \) and \( \mu_2 \) are selected to meet the prescribed Type-I error constraints. 
Given that $p_0 \leq p_1$, and $c_0 \leq c_1$, the tests described by \eqref{eq:np_enc1} and \eqref{eq:np_enc2}  are equivalent, respectively, to the conditions:
\begin{align}
    w(\mathbf{x}^n)  &< \lambda_1, \\
     w(\mathbf{y}^n)  &< \lambda_2.
\end{align}
$\lambda_1 \text{ and } \lambda_2 \in \mathbb{N}$ are  threshold values chosen so as to satisfy the constraints $\alpha_n^{(1)} <  \epsilon$, and $\alpha_n^{(2)} <  \epsilon$, respectively. 

The decision function $g^{(n)}:\{0,1\}\times \{0,1\} \to \{0,1\}$ is described as follows. Let us denote \( b_1 = f_1^{(n)}\left(\mathbf{x}^{n}\right) \) and \( b_2 = f_2^{(n)}\left(\mathbf{y}^{n}\right) \) as the  1-bit produced by Encoder 1 and Encoder 2, respectively. Upon receiving \( b_1 \) and \( b_2 \), the decoder decides that \( g^{(n)}\left(b_1, b_2\right) = i \) if \( b_1 = i \) and \( b_2 = i \) for \( i = 0, 1 \). Otherwise, the decoder relies on the decision of Encoder 1.  This choice is motivated by the fact that according to the model defined in~\eqref{binary_dht},  $Y$ is a noisy version of $X$, so that $\mathbb{P} (Y =1) > \mathbb{P} (X =1)$ under both hypotheses $\mathcal{H}_0$ and $\mathcal{H}_1$. Note that other strategies may be considered depending on the values of $p_0,p_1,c_0,c_1$. 

From the NP lemma, we can show that the Type-I and Type-II error probabilities 
of the separate scheme are given by 
\begin{align}
\alpha_{n}^{(s)} &= 1 - \sum_{k=0}^{\gamma_1} \binom{n}{k} p_0^k (1 - p_0)^{n - k}, 
\label{alpha_sep}\\
\beta_n^{(s)} &= \sum_{k=0}^{\gamma_1} \binom{n}{k} p_1^k (1 - p_1)^{n - k},
\label{beta_sep}
\end{align}
where the exponents $(s)$ refer to the separate scheme. 
These expressions are also applicable to the asymmetric setup, where  \( Y \) serves as side information at the decoder.

\subsection{Truncation scheme}\label{sec:trunc_sym}
The truncation scheme consists of sending the first $l$ symbols of the source vector $\mathbf{x}^n$ and $\mathbf{y}^n$ at the coding rate $R_1 = R_2 = l/n$ at the decoder\footnote{Considering  $R_1 \neq R_2$ would result in sending a different number of symbols from  $\mathbf{x}^n$ and $\mathbf{y}^n$. Given that the sources are i.i.d., additional symbols from one source may not be efficiently exploited by an NP test accounting for the correlation between the two sources. This is why we always consider $R_1=R_2$ in our analyses. }. The decoder can then perform a standard NP test~\cite{lehmann2005testing} on the pair $(\mathbf{x}^l,\mathbf{y}^l)$. Under a certain constraint $\alpha_n^{(t)} <  \epsilon$ on Type-I error probability for the truncation scheme, the NP lemma~\cite{lehmann2005testing} states that the following test$:$
\begin{equation}\label{eq:NPtest}
\mathbb{P}_{1}(\mathbf{x}^l,\mathbf{y}^l) < \mu  \mathbb{P}_{0}(\mathbf{x}^l,\mathbf{y}^l) ,
\end{equation}
  where $\mathcal{H}_0$ is decided if the inequality is satisfied,
minimizes Type-II error probability $\beta_n^{(t)}$. In~\eqref{eq:NPtest}, $\mu$ is a threshold value chosen to satisfy the Type-I error constraint. Given that $p_0 \leq p_1$ and $c_0 \leq c_1$, the previous expression~\eqref{eq:NPtest} 
 simplifies to
 \begin{equation}
     w(\mathbf{x}^l) \log\frac{p_1(1-p_0)}{p_0(1-p_1)} + w(\mathbf{z}^l) \log\frac{c_1(1-c_0)}{c_0(1-c_1)}  <\tau_t,
\label{eq:NP_criteria_tronc_sym}
 \end{equation}
where $\mathbf{z}^l = \mathbf{x}^l \oplus \mathbf{y}^l$, and $\tau_t$ is a threshold value chosen so as to satisfy the Type-I error constraint. Note that during the derivation from~\eqref{eq:NPtest} to~\eqref{eq:NP_criteria_tronc_sym}, a relationship between $\mu$ and $\tau_l$ appears. However, in hypothesis testing, it suffices to set the threshold $\tau_l$ so as to satisfy the Type-I error probability constraint. Therefore, the expression of $\tau_l$ with respect to $\mu$ does not influence the final expression of the test. This is why we do not provide this expression here.  

From the NP lemma, given that $p_0 < p_1$ and  $c_0 < c_1$, the analytical expressions of Type-I and Type-II errors of the truncation scheme are given by
\begin{align}
    \alpha_n^{(t)} &= \sum_{\substack{(\lambda, j) :\\
    T_{\lambda,j}\geq\tau_t}}  \binom{l}{\lambda} p_0^\lambda (1 - p_0)^{l - \lambda} \binom{l}{j} c_0^j (1 - c_0)^{l - j} 
\label{eq:alpha_tronc_sym} \\
\beta_n^{(t)} &= \sum_{\substack{(\lambda, j) :\\
T_{\lambda,j}\leq\tau_t}}  \binom{l}{\lambda} p_1^\lambda (1 - p_1)^{l - \lambda} \binom{l}{j} c_1^j (1 - c_1)^{l - j}
\label{eq:beta_tronc_sym}
\end{align}
where the exponent $(t)$ refers to the truncation scheme, and $T_{\lambda,j} =  \lambda \log_2\frac{p_1(1-p_0)}{p_0(1-p_1)} + j\log_2\frac{c_1(1-c_0)}{c_0(1-c_1)}$.



We will compare the performance of the separate scheme
and the truncation scheme in terms of Type-I and Type-II
error probabilities in the numerical results section. We will also compare the performance of the  truncation scheme with the quantization and quantize-binning schemes we propose in the following sections.

\section{Quantization scheme}\label{sec:quant}
In their seminal work \cite{ahlswede1986hypothesis}, Ahlswede and Csiszar introduced the first Information-Theoretic DHT scheme based only on a quantizer. 
Here, we present a practical implementation of this scheme for short-length sequences by utilizing linear block codes.
\subsection{Code construction for the symmetric setup}\label{sec:code_quant}
To practically implement binary quantization for the symmetric setup, we follow the approach of\cite{fridrich2007binary,wainwright2009low,castanheira2010lossy} and consider a generator matrix $G_q$ with dimension $n\times k$ of a linear block code. According to the ML rule, for given source sequences $\mathbf{x}^n$ and $\mathbf{y}^n$, the encoders produce vectors $\mathbf{u}_{q}^{k}$ and $\mathbf{t}_{q}^{k}$ of length $k$ bits as~\cite{chandar2006information}
\begin{align}
 \mathbf{u}_{q}^k& =\arg \min _{\mathbf{u}^k} d\left(G_q \mathbf{u}^k, \mathbf{x}^n\right)  \label{quant_x}, \\
\mathbf{t}_{q}^k &=\arg \min _{\mathbf{t}^k} d\left(G_q \mathbf{t}^k, \mathbf{y}^n\right).
    \label{quant_y}
\end{align}
The vectors $\mathbf{u}_{q}^{k}$ and $\mathbf{t}_{q}^{k}$ are the compressed versions of $\mathbf{x}^n$ and $\mathbf{y}^n$, respectively. 
We further denote $\mathbf{x}_{q}^n = G_q \mathbf{u}_{q}^k$, and $\mathbf{y}_{q}^n = G_q \mathbf{t}_{q}^k$. 

In~\eqref{quant_x}, and ~\eqref{quant_y}, the key difficulty lies in determining the quantized vectors $\mathbf{u}_{q}^k $ and $\mathbf{t}_{q}^k$ that achieve the minimum Hamming distance. In~\cite{fridrich2007binary,wainwright2009low,castanheira2010lossy}, the matrix $G_q$ is constructed as an LDGM codes, which enables the use of a low complexity message-passing algorithm called Bias-Propagation to solve~\eqref{quant_x},~\eqref{quant_y}. The schemes introduced in~\cite{fridrich2007binary,wainwright2009low,castanheira2010lossy} consider long codes (more than $10^3$ bits). But at short length, LDGM codes are penalized by their low minimum distances between codewords. Instead, we opt to consider well known short linear block codes such as BCH and Reed-Muller codes, since they have good minimum distance properties. Unfortunately, their generator matrices are not sparse, which prevents the use of  the Bias-Propagation algorithm.

To find $\mathbf{u}_q^k$ and $\mathbf{t}_q^k$ 
we reformulate problems~\eqref{quant_x} and~\eqref{quant_y} as \begin{align}\label{quant_x_eq}
\mathbf{x}_{q}^n & = \arg\min_{\mathbf{x}_{q}^n} d(\mathbf{x}_{q}^n, \mathbf{x}^n) ~ \text{s.t.} ~ H_q \mathbf{x}_q^n = \mathbf{0}^m, \\ \label{quant_y_eq}
\mathbf{y}_{q}^n & = \arg\min_{\mathbf{y}_{q}^n} d(\mathbf{y}_{q}^n, \mathbf{y}^n) ~ \text{s.t.} ~ H_q \mathbf{y}_q^n = \mathbf{0}^m,
\end{align}
where $H_q$ is a parity check matrix of size $m\times n$ of the code defined by $G_q$, with $m=n-k$. 
Once the closest valid codewords $\mathbf{x}_{q}^n$ and $\mathbf{y}_q^n$ are identified, the corresponding compressed vectors $\mathbf{u}_{q}^k$ and $\mathbf{t}_{q}^k$ can be retrieved by solving the linear systems $G_q \mathbf{u}_{q}^k = \mathbf{x}_{q}^n$ and $G_q \mathbf{u}_{q}^k = \mathbf{x}_{q}^n$, typically via Gaussian elimination.


Problems~\eqref{quant_x_eq} and~\eqref{quant_y_eq}, which are equivalent to~\eqref{quant_x} and~\eqref{quant_y}, respectively, are a standard channel coding formulation. In this sense, a natural approach might be to apply Belief Propagation (BP) decoding~\cite{richardson2008modern} to~\eqref{quant_x_eq} and~\eqref{quant_y_eq}, as commonly done for LDPC codes in channel coding. However, BP decoders perform poorly for binary quantization, especially at short block lengths. This arises because, unlike in channel decoding where the received vector is a noisy version of a transmitted codeword, in binary quantization the source vectors $\mathbf{x}^n$ and $\mathbf{y}^n$ are typically far from any valid codeword. Consequently, BP often fails to converge to a valid solution in this context.


In this work, we do not rely on BP decoders, which are not well-suited to short codes or to binary quantization. Instead, we propose solving problems~\eqref{quant_x_eq} and~\eqref{quant_y_eq} exactly using Maximum Likelihood (ML) decoders tailored to the considered codes. Specifically, we use generic syndrome-based decoders~\cite[Section 3.1.4]{ryan2009channel}, which can be applied to a broad range of short block codes, including BCH and Reed-Muller codes. When applicable, code-specific decoders such as the Berlekamp-Massey algorithm for BCH codes~\cite[Section 3.2.2]{ryan2009channel} may also be employed to further reduce complexity or improve performance.



\subsection{Hypothesis test for the symmetric setup}
The codewords $\mathbf{u}_{q}^k$ and $\mathbf{t}_{q}^k$ are transmitted to the decoder at code rates $R_1 = R_2 = k/n$. The decoder first reconstructs the quantized vectors  $\mathbf{x}_{q}^n = G_q\mathbf{u}_{q}^k$ and $\mathbf{y}_{q}^n = G_q\mathbf{t}_{q}^k $. It then applies a NP test to the pair $(\mathbf{x}_{q}^n,\mathbf{y}_{q}^n)$. Under a certain constraint $\alpha_n^{(q)} <  \epsilon$ on Type-I error probability for the quantization scheme, the NP lemma~\cite{lehmann2005testing} states that the following test$:$
\begin{equation} \mathbb{P}_1\left(\mathbf{x}_q^n, \mathbf{y}_q^n\right) \leq\mu_q\mathbb{P}_0\left(\mathbf{x}_q^n, \mathbf{y}_q^n\right),
\label{eq:NP_test_quant_sym}
 \end{equation}
 where $\mathcal{H}_0$ is decided if the inequality is satisfied, 
minimizes Type-II error probability $\beta_n^{(q)}$. In~\eqref{eq:NP_test_quant_sym}, $\mu_q$ is a threshold chosen so as to satisfy the Type-I error constraint. Interestingly, the entropy-check condition~\eqref{eq:empirical_entropy} from the Information-Theoretic analysis only involved the probability distribution under $\mathcal{H}_0$, while the NP test~\eqref{eq:NP_test_quant_sym} compares the distributions under the two hypothesis. 
Moreover, while condition~\eqref{eq:empirical_entropy} arises in an asymptotic setting, the NP test remains optimal for any finite $n$.

However, computing the joint distributions  $\mathbb{P}_0 (\mathbf{x}_q^n, \mathbf{y}_q^n)$ and $\mathbb{P}_1 (\mathbf{x}_q^n, \mathbf{y}_q^n)$ in \eqref{eq:NP_test_quant_sym} is non-trivial, as the underlying code $G_q$ introduces statistical dependencies within the vectors $\mathbf{x}_q^n$ and $\mathbf{y}_q^n$. To simplify the analysis, we model the vectors \( \mathbf{x}_q^n \) and \( \mathbf{v}_q^n = \mathbf{x}_q^n \oplus \mathbf{y}_q^n \) as realizations of i.i.d. Bernoulli random variables: \( X_q \sim \text{Bern}(\hat{p}) \) with $\hat{p} = \hat{p}_0$ under $\mathcal{H}_0$ and $\hat{p} = \hat{p}_1$ under $\mathcal{H}_1$, and \( V_q \sim \text{Bern}(\hat{c}) \) with $\hat{c} = \hat{c}_0$ under $\mathcal{H}_0$ and $\hat{c} = \hat{c}_1$ under $\mathcal{H}_1$. 
 

Under these assumptions, the test~\eqref{eq:NP_test_quant_sym} can be reformulated as
 \begin{align}
w(\mathbf{x}^{n}_{q}) \log_2\frac{\hat{p}_1(1-\hat{p}_0)}{\hat{p}_0(1-\hat{p}_1)} +  w(\mathbf{v}^{n}_{q}) \log_2\frac{\hat{c}_1(1-\hat{c}_0)}{\hat{c}_0(1-\hat{c}_1)}  \leq \tau_q,
\label{eq:NP_criteria_quant_sym}
 \end{align}
where $\tau_q$ is chosen to meet the Type-I error probability requirement. 
The parameters \(( \hat{p}_0, \hat{c}_0\)) and \(( \hat{p}_1, \hat{c}_1\)) are estimated through Monte-Carlo simulations. Althoug $\mathbf{x}_q^n$ and $\mathbf{v}_q^n$ are not strictly i.i.d. in practice,
the numerical results presented later on in the paper demonstrate that the proposed test provides accurate decisions under this approximation.


 \subsection{Code construction and hypothesis test for the asymmetric setup}\label{sec:code_quant_assy}
 In the asymmetric setup, the coding scheme is obtained in a straightforward manner from the symmetric setup. In this case, only $\mathbf{x}^n$ is quantized and transmitted at rate $R_1 = k/n$, while $\mathbf{y^n}$ is available at the decoder.
 Therefore, the decoder first computes the quantized vector  $\mathbf{x}_q^n = G_q \mathbf{u}_{q}^k$. Then the NP test remains the same as~\eqref{eq:NP_criteria_quant_sym}, except that $\mathbf{y}_q^n$ is replaced by $\mathbf{y}^n$ in the expression of $\mathbf{v}_q^n$.  
 
 Note that in the special case where $p_0 = p_1$, the NP test~\eqref{eq:NP_test_quant_sym} reduces to
\begin{equation}
 \sum_{i=1}^n (x_{q,i} \oplus y_i)  < \lambda_q,
\label{eq:NP_criteria_quant_assym}
\end{equation} 
where $\lambda_q$ is the threshold chosen so as to satisfy the Type-I error probability constraint.

\subsection{Theoretical analysis of the quantization scheme}\label{sec:alpha_beta_theo_quant}


In this section, we provide a theoretical analysis of the Type-I and Type-II error probabilities for the proposed practical quantization scheme under the asymmetric setup, focusing on the special case where $p_0 = p_1 = 1/2$.  We derive exact analytical expressions for the error probabilities of the scheme considering the generator matrix $G_q$.  Extensions to the symmetric setup and to cases where $p_0 \neq p_1$ involve significantly more complex analyses and are therefore left for future work. 

We begin by introducing the notation associated with the code defined by the generator matrix $G_q$. Throughout the analysis, we assume that the all-zero codeword $\mathbf{x}_q^n$ is transmitted. 
 Due to symmetry, the quantization error probability is independent of the transmitted codeword~\cite{richardson2008modern}. 

 Let $\mathcal{C}_0^{(q)}$ denote the decision region, or coset, corresponding to $\mathbf{x}_q^n = \mathbf{0}^n$, defined as
\begin{equation}\notag
\mathcal{C}_0^{(q)} := \left\{ \mathbf{x}^n \in \{0,1\}^n : \arg \min_{\mathbf{u}^k}  d(G_q\mathbf{u}^k,\mathbf{x}^n) = \mathbf{0}^k \right\}. 
\end{equation}
In other words, $\mathbf{x}^n \in \mathcal{C}_0^{(q)}$ implies that that the solution of~\eqref{quant_x_eq} for $\mathbf{x}_q^n$ is $\mathbf{0}^n$.
Next consider the set of integers $\{E_\gamma^{(q)}\}_{\gamma \in \llbracket 0,d_{\text{max}}^{(q)}\rrbracket}$ where $E_\gamma^{(q)}$ is the number of sequences $\mathbf{x}^n$ of Hamming weight $\gamma$, also referred to as number of coset leaders of weight $\gamma$, that belong to the coset $\mathcal{C}_0^{(q)}$. 
In addition, $d_{\text{max}}^{(q)}$ is the maximum possible weight in $\mathcal{C}_0^{(q)}$. 
The total number of coset leaders is then given by $ N_0^{(q)} = \sum_{\gamma=0}^{d_{\text{max}}^{(q)}} E_{\gamma}^{(q)}$.  The reader is referred to~\cite[Section 3.1.4]{ryan2009channel} for more details about these concepts which were originally introduced in the context of channel coding. 

 With these definitions in place, we proceed to derive the theoretical expressions for the Type-I and Type-II error probabilities of the proposed quantization scheme under the asymmetric setup.
\begin{prop}\label{prop:error_quantization}
Consider the quantization scheme in the asymmetric setup, with $p_0 = p_1 = 0.5$,  and a fixed threshold value $\lambda_q$. Type-I and Type-II error probabilities of this scheme are given by
\begin{align}
  \alpha_n^{(q)} & = 1 - \frac{1}{N_0^{(q)}} \sum_{\lambda=0}^{\lambda_q} \sum_{\gamma = 0}^{d_{\text{max}}^{(q)}} \sum_{j=0}^n E_\gamma^{(q)} \Delta_{\lambda,j,\gamma} \binom{n}{j} c_0^j (1-c_0)^{n-j}, \label{eq:alphaq}  \\
 \beta_n^{(q)} & =  \frac{1}{N_0^{(q)}} \sum_{\lambda=0}^{\lambda_q} \sum_{\gamma = 0}^{d_{\text{max}}^{(q)}} \sum_{j=0}^n E_\gamma^{(q)} \Delta_{\lambda,j,\gamma} \binom{n}{j} c_1^j (1-c_1)^{n-j}, \label{eq:betaq}
\end{align}
where
\begin{equation}
\label{Delta_def}
\Delta_{\lambda,j,\gamma} = \frac{\Gamma_{\lambda,j,\gamma}}{\sum_{i=0}^{\text{max}(\lambda, j)}\binom{\lambda}{i} \binom{n-\lambda}{j - i} },
\end{equation}
and, for $\gamma = j + \lambda - 2u$ and $0 \leq u \leq \min(\lambda, j) \leq n$,
\begin{equation}\label{eq:Gamma_def}
 \Gamma_{\lambda,j,\gamma} = \binom{\lambda}{u} \binom{n-\lambda}{j - u} . 
\end{equation}
\end{prop}
\begin{proof} 
By symmetry due to the linear block code, the quantizer error probability is independent of the transmitted codeword~\cite{richardson2008modern}. Therefore, it is sufficient to consider the all-zero codeword $\mathbf{x}_q^n = \mathbf{0}^n$. From \eqref{eq:NP_criteria_quant_assym}, we develop
\begin{align}\label{eq:proof_q1}
   \alpha_n^{(q)} & = 1 - \sum_{\lambda=0}^{\lambda_q} \mathbb{P}_0(w(\mathbf{Y}^n)  = \lambda) \\ \notag
   & = 1 -   \sum_{\lambda=0}^{\lambda_q}\sum_{\gamma=0}^{d_{\text{max}}^{(q)}} \frac{E_\gamma^{(q)}}{N_0^{(q)}} \mathbb{P}_0(w(\mathbf{Y}^n)=\lambda | w(\mathbf{X}^n) = \gamma) \\ \notag
   & = 1 -   \sum_{\lambda=0}^{\lambda_q}\sum_{\gamma=0}^{d_{\text{max}}^{(q)}} \frac{E_\gamma^{(q)}}{N_0^{(q)}} \sum_{j=0}^n \mathbb{P}_0(d(\mathbf{X}^n,\mathbf{Y}^n) = j ) \Delta_{\lambda,j,\gamma} \\ \notag
   & = 1 - \sum_{\lambda=0}^{\lambda_q}\sum_{\gamma=0}^{d_{\text{max}}^{(q)}} \frac{E_\gamma^{(q)}}{N_0^{(q)}} \sum_{j=0}^n \binom{n}{j} c_0^j (1-c_0)^{n-j} \Delta_{\lambda,j,\gamma}.
 \end{align}
This leads to~\eqref{eq:alphaq}. 
To obtain~\eqref{eq:betaq}, we remark that $ \beta_n^{(q)} = \sum_{\lambda=0}^{\lambda_q} \mathbb{P}_1(w(\mathbf{Y}^n)  = \lambda)$. Following the same steps as in~\eqref{eq:proof_q1}, and by replacing $c_0$ by $c_1$, the proof is completed. 
$~~~~~~~~~~~~~~\square$
\end{proof}

To the best of the authors’ knowledge, these theoretical results are novel and differ from both the classical Information-Theoretic analysis of DHT and existing results in channel coding. 
 Especially, while error probability expressions are well-established for linear block codes in the context of channel coding, such analytical characterizations had not been derived for the DHT problem, where the threshold parameter $\lambda_q$ impacts both Type-I and Type-II error probabilities. 
 These new analytical expressions enable the evaluation of decision performance without relying on computationally intensive Monte Carlo simulations. Consequently, they facilitate code design by only considering the code parameters such as $E_{\gamma}^{(q)}, N_0^{(q)}, \text{ and }  d_{\text{max}}^{(q)}$, as demonstrated in~\cite{fatemeh2024quantize}. 


\section{Quantize-binning scheme}
\label{sec:quant_bin}
In their seminal work \cite{shimokawa1994error}, Shimokawa et al. introduced the quantize-binning scheme for DHT. This scheme leverages the correlation between the sources $X$ and $Y$ to reduce the code rate after compression. We now introduce a practical short-length implementation of this scheme by using linear block codes.
\subsection{Code construction for the symmetric setup}
To practically implement the quantize-binning scheme, we again consider the generator matrix $G_q$ of size $n\times k$ of a linear block code. Additionally, we consider the parity-check matrix $H_b$ of size $\ell \times k$ of \textit{another} linear block code. In the symmetric setup, given the source vectors $\mathbf{x}^n$ and $\mathbf{y}^n$, the encoders apply the quantization method described by~\eqref{quant_x} and~\eqref{quant_y} to obtain the sequences $ \mathbf{u}_{q}^{k}$ and $ \mathbf{t}_{q}^{k}$, respectively. Then, the encoders use the parity check matrix $H_b$ to compute the syndromes 
\begin{equation}
    \mathbf{r}^{\ell}=H_b  \mathbf{u}_{q}^{k},
    \label{eq:syndrom_bin_x}
\end{equation}
and
\begin{equation}
    \mathbf{s}^{\ell}=H_b  \mathbf{t}_{q}^{k},
    \label{eq:syndrom_bin_y}
\end{equation}
both of length $\ell$. 
This follows the same approach as in~\cite{xiong2004distributed, liveris2002compression, savard2013improved,ye2019optimized},  where binning is performed using the parity-check matrix of an LDPC code.
In contrast, and consistent with our earlier discussion, we employ efficient short-length linear block codes such as BCH and Reed–Muller codes.

\subsection{Hypothesis test for the symmetric setup}
The syndromes  $\mathbf{r}^{\ell}$ and $\mathbf{s}^{\ell}$ are then transmitted to the decoder at rates $R_1 = R_2 = \ell/n$. At the decoder, as discussed in Section \ref{sec:quant}, we avoid using message-passing algorithms such as BP decoders since they do not perform well with short-length codes. Instead, we directly consider the ML rule. Therefore, the decoder first  identifies by exhaustive search, vectors  $\hat{\mathbf{u}}_{q}^k$ and $\hat{\mathbf{t}}_{q}^k$ as
\begin{align}
   ( \hat{\mathbf{u}}_{q}^k, \hat{\mathbf{t}}_{q}^k) & =\arg \min _{\mathbf{u}^{k}, \mathbf{t}^{k}} d\left(G_q \mathbf{u}^{k}, G_q \mathbf{t}^{k}\right) \text{ s.t. } \\ \notag
    &H_b\mathbf{u}^k = \mathbf{r}^{\ell}, \text{ and } H_b\mathbf{t}^m = \mathbf{s}^{\ell}.
    \label{eq:decoder_qandb_sym}
\end{align}
We then compute $\mathbf{\hat{x}}_{q}^n = G_q\hat{\mathbf{u}}_{q}^k$ and $\mathbf{\hat{y}}_{q}^n = G_q\hat{\mathbf{t}}_{q}^k $, and, apply the following NP test
\begin{equation} \mathbb{P}_1\left(\mathbf{\hat{x}}_{q}^n, \mathbf{\hat{y}}_{q}^n\right)\leq\mu_{q,b} \mathbb{P}_0\left(\mathbf{\hat{x}}_{q}^n, \mathbf{\hat{y}}_{q}^n\right),
\label{eq:NP_quantb_sym}
\end{equation}
where $\mathcal{H}_0$ is decided if the inequality is satisfied. In the previous expression, $\mu_{q,b}$ is a threshold chosen so as to satisfy the Type-I error constraint.

As in Section \ref{sec:code_quant}, we note that computing the joint distributions $\mathbb{P}_0(\mathbf{\hat{x}}_{q}^n, \mathbf{\hat{y}}_{q}^n)$ and $\mathbb{P}_1(\mathbf{\hat{x}}_{q}^n, \mathbf{\hat{y}}_{q}^n)$ in \eqref{eq:NP_quantb_sym} is difficult. Therefore, we adopt the same assumptions as in Section \ref{sec:code_quant}. Specifically, we assume that \( \mathbf{\hat{x}}_{q}^n \) and \( \mathbf{\hat{v}}_{q}^n = \mathbf{\hat{x}}_{q}^n \oplus \mathbf{\hat{y}}_{q}^n \) are the realizations of i.i.d. random variables \( \hat{X}_{q}\sim \text{Bern}(\hat{p}_b) \) (with $\hat{p}_b = \hat{p}_{0,b}$ under $\mathcal{H}_0$, and $\hat{p}_b = \hat{p}_{1,b}$, under $\mathcal{H}_1$),  and \( \hat{V}_{q} \sim \text{Bern}(\hat{c}_b) \) (with $\hat{c}_b = \hat{c}_{0,b}$ under $\mathcal{H}_0$, and $\hat{c}_b = \hat{c}_{1,b}$, under $\mathcal{H}_1$). 
The NP test~\eqref{eq:NP_quantb_sym} can then be rewritten as
\begin{align}
     w(\mathbf{\hat{x}}^{n}_{q}) \log_2\frac{\hat{p}_{1,b}(1-\hat{p}_{0,b})}{\hat{p}_{0,b}(1-\hat{p}_{1,b})} +  w(\mathbf{\hat{v}}^{n}_{q}) \log_2\frac{\hat{c}_{0,b}(1-\hat{c}_{0,b})}{\hat{c}_{0,b}(1-\hat{c}_{1,b})}  \leq \tau_{q,b},
\label{eq:NP_criteria_quantb_sym}
 \end{align} 
 where $\tau_{q,b}$ is a threshold chosen so as to satisfy the Type-I error probability. 
 The values $\hat{p}_{0,b}$, $\hat{c}_{0,b}$, $\hat{p}_{1,b}$, and $\hat{c}_{1,b}$ are estimated through Monte-Carlo simulations.

 \subsection{Code construction and hypothesis test for the asymmetric setup} \label{sec:compare_inf_quant}
 In the asymmetric case, only $\mathbf{x}^n$ is quantized and binned into $\mathbf{r}^{\ell}$, according to~\eqref{eq:syndrom_bin_x}. The vector $\mathbf{u}^{\ell}$  is then transmitted at rate $R_1 = \ell/n$, while $\mathbf{y}^n$ serves as side information at the decoder. At the receiver, the vector $\hat{\mathbf{u}}^k$ is identified by solving the ML rule
\begin{equation}\label{eq:decoder_qandb}
 \hat{\mathbf{u}}_q^k = \arg\min_{\mathbf{u}^k} ~ d(G_q\mathbf{u}^k,\mathbf{y}^n) \text{ s.t. } H_b\mathbf{u}^k = \mathbf{r}^{\ell} .
\end{equation}
through an exhaustive search. 
Next, in the asymmetric setup, the NP test is the same as~\eqref{eq:NP_quantb_sym}, except that $\mathbf{\hat{y}}_{q}^n$ is replaced by $\mathbf{y}^n$ in the derivation. In the special case where $p_0 = p_1$, this test reduces to 
\begin{equation}
 \sum_{i=1}^n (\hat{x}_{q,i} \oplus y_i)  < \lambda_{qb},
 \label{eq:NP_criteria_quantb_asym}
\end{equation}
where $\hat{\mathbf{x}}_{q}^n = G_q \hat{\mathbf{u}}^k$, and  $\lambda_{qb}$ is a threshold chosen so as to satisfy the Type-I error probability.


\subsection{Theoretical analysis of the quantize-binning scheme}\label{sec:alpha_beta_theo_quantb}
We now provide exact analytical expressions for the Type-I
and Type-II error probabilities of the quantize-binning scheme, under the asymmetric setup and in the particular case where $p_0 = p_1 = 1/2$. Extensions to the symmetric setup and the general case where $p_0 \neq p_1$ are considerably more complex and are therefore left for future work. 

As in Section~\ref{sec:alpha_beta_theo_quant}, we assume that the all-zero codeword is transmitted.  We use the definitions introduced in Section~\ref{sec:alpha_beta_theo_quant} for $\mathcal{C}_0^{(q)}$ and $E_{\gamma}^{(q)}$ associated with the generator matrix $G_q$. Additional definitions are necessary for the quantize-binning scheme, which corresponds to the concatenation of the two codes used for quantization and binning.

We first define the decision region, or coset, $\mathcal{C}_0^{(qb)}$ as
\begin{equation}\label{eq:decision_region_qb}
\mathcal{C}_0^{(qb)} := \left\{ \mathbf{y}^n \in \{0,1\}^n : \arg \min_{\mathbf{u}^k}  d(G_q\mathbf{u}^k,\mathbf{y}^n) = \mathbf{0}^k \right\}, 
\end{equation}
for the all-zero codeword of the quantize-binning scheme.
Importantly, the condition  $H_b \mathbf{u}^k = \mathbf{r}^{\ell}$ on the syndrome is not explicitly included in~\eqref{eq:decision_region_qb}, as it is automatically satisfied for $\mathbf{u}^k = \mathbf{0}^k$ due to the linearity of the code. Especially, a side information vector $\mathbf{y}^n$ belongs to $ \mathcal{C}_0^{(qb)}$ if the solution of~\eqref{eq:decoder_qandb} for this vector is $\hat{\mathbf{u}}_q^k = \mathbf{0}^k$. 

Next, we define the set of integers $\{E_\nu^{(qb)}\}_{\nu \in \llbracket 0,d_{\text{max}}^{(qb)}\rrbracket}$, where $E_\nu^{(qb)}$ denotes the number of sequences $\mathbf{y}^n$ with a Hamming weight $\nu$ that belong to the decision region $\mathcal{C}_0^{(qb)}$. These quantities are also referred to as the number of coset leaders of weight $\nu$. Here, $d_{\text{max}}^{(qb)}$ denotes the maximum possible weight in $\mathcal{C}_0^{(qb)}$. Additionally, we define the set of integers $\{A_t^{(qb)}\}_{t \in \llbracket 0,n \rrbracket }$, where $A_t^{(qb)}$ denotes the number of sequences $\mathbf{x}_q^n$ with a Hamming weight~$t$ that can be expressed as $\mathbf{x}_q^n = G_q \mathbf{u}_q^k$, for some $\mathbf{u}_q^k$ satisfying $H_b \mathbf{u}_q^k = \mathbf{0}^{\ell}$. Thus, the set  $\{A_t^{(qb)}\}_{t \in \llbracket 0,n \rrbracket }$ corresponds to the code weight distribution~\cite[Section 3.1.3]{ryan2009channel} of the concatenated code.
\begin{prop}\label{prop:qandb}
 For the quantize-binning scheme considering $p_0 = p_1 = 1/2$, and for a fixed threshold value $\lambda_{qb}$, Type-I and Type-II error probabilities are given by
 \begin{align}
  \alpha_n^{(qb)} & = 1 - \mathbb{P}_{B}(c_0) -  \mathbb{P}_{\bar{B}}(c_0),\\
  \beta_n^{(qb)} & = \mathbb{P}_{B}(c_1) +  \mathbb{P}_{\bar{B}}(c_1),
  \label{eq:beta_qb}
 \end{align}
 where
 \begin{align}\label{eq:Pb0Blambda}
  \mathbb{P}_{B}(\delta)  & = \sum_{\nu=0}^{\min(d_{\max}^{(qb)},\lambda_{qb})} \frac{E_{\nu}^{(qb)}}{\binom{n}{\nu}} \sum_{\gamma=0}^{d_{\max}^{(q)}} \frac{E_\gamma^{(q)}}{N_0^{(q)}} \sum_{j=0}^n \Gamma_{\nu,j,\gamma} \delta^j (1-\delta)^{n-j} ,  \\   \label{eq:Pb0barBlambda}
  \mathbb{P}_{\bar{B}}(\delta) & = \sum_{i=0}^n \left[ \left( \sum_{\gamma=0}^{d_{\text{max}}^{(q)}} \frac{E_{\gamma}^{(q)}}{N_0^{(q)}} \sum_{j=0}^n \Gamma_{i,j,\gamma} \delta^j (1-\delta)^{n-j} \right) \right. \\ \notag
  & ~~~~~~~~~~~~~~~~~~~~~~ \left. \times \left( \sum_{t=1}^n \sum_{\nu=0}^{\lambda_{qb}} \frac{E_{\nu}^{(qb)}}{\binom{n}{\nu}} \frac{A_t^{(qb)} \Gamma_{i,\nu,t}}{\binom{n}{i}}  \right) \right] .
\end{align}
\end{prop}

\begin{proof}
We consider the all-zero codeword $\mathbf{x}_q^n = \mathbf{0}$. 
 Under the hypothesis $\mathcal{H}_0$, we express
 \begin{equation}
  \alpha_n^{(qb)} = 1 -  \mathbb{P}_0(\mathcal{\widehat{H}}_0,B) -  \mathbb{P}_0(\mathcal{\widehat{H}}_0,\bar{B}),
 \end{equation}
where $B$ is the event that the correct sequence $\hat{\mathbf{x}}_q^n = \mathbf{x}_q^n$ is retrieved by the decoder, while $\bar{B}$ is the event that an incorrect sequence $\mathbf{\hat{x}}_q^n \neq \mathbf{x}_q^n $ is output by the decoder. In addition, $\mathcal{\widehat{H}}_0$ is the event that the hypothesis $\mathcal{H}_0$ is chosen by the decoder. We further denote  $ \mathbb{P}_{B}(p_0) =  \mathbb{P}_0(\mathcal{\widehat{H}}_0,B)$ and $ \mathbb{P}_{\bar{B}}(p_0) =  \mathbb{P}_0(\mathcal{\widehat{H}}_0,\bar{B})$. 
We then express
\begin{align}
\mathbb{P}_{B}(p_0) &=  \sum_{\nu=0}^n \mathbb{P}_0(w(\mathbf{Y}^n)=\nu) \mathbb{P}_0(\mathcal{\widehat{H}}_0,B|w(\mathbf{Y}^n)=\nu) \\ 
& = \sum_{\nu=0}^{\min(d_{\max}^{(qb)},\lambda_{qb})}  \mathbb{P}_0(w(\mathbf{Y}^n)=\nu) \frac{E_{\nu}^{(qb)}}{\binom{n}{\nu}} .
\end{align}
By following the same steps as in the proof of Proposition~\ref{prop:error_quantization}, we can show that \begin{equation}\label{eq:Pwy}
\mathbb{P}_0(w(\mathbf{Y}^n)=\nu) =  \sum_{\gamma=0}^{d_{\max}^{(q)}} \frac{E_\gamma^{(q)}}{N_0^{(q)}} \sum_{j=0}^n \Gamma_{\nu,j,\gamma} c_0^j (1-c_0)^{n-j} ,
\end{equation}
which provides~\eqref{eq:Pb0Blambda}. 
We then write 
\begin{equation}
  \mathbb{P}_{\bar{B}}(p_0) = \sum_{i=0}^n \mathbb{P}_0(w(\mathbf{Y}^n)=i) \mathbb{P}_0(\mathcal{\widehat{H}}_0,\bar{B}|w(\mathbf{Y}^n)=i),
\end{equation}
where $\mathbb{P}_0(w(\mathbf{Y}^n)=i)$ is given by~\eqref{eq:Pwy}. We develop
\begin{align}\notag
 & \mathbb{P}_0(\mathcal{\widehat{H}}_0,\bar{B}|w(\mathbf{Y}^n)=i)  \\ & = \sum_{t=1}^n \sum_{\nu=0}^{\lambda_{qb}} \mathbb{P}_0(w(\hat{\mathbf{X}}_q^n) = t,d(\hat{\mathbf{X}}_q^n,\mathbf{Y}^n) = \nu |w(\mathbf{Y}^n) = i) \\
 & = \sum_{t=1}^n \sum_{\nu=0}^{\lambda_{qb}} \frac{E_\nu^{(qb)}}{\binom{n}{\nu}}  \frac{A_t^{(qb)} \Gamma_{i,\nu,t}}{\binom{n}{i}}.
\end{align}
This provides the expression of $ P_{\bar{B}}(c_0)$ in~\eqref{eq:Pb0barBlambda}.

We can derive the expression of $\beta_n^{(qb)}$ in~\eqref{eq:beta_qb} by noticing that $\beta_n^{(qb)} = \mathbb{P}_{B}(c_1) + \mathbb{P}_{\bar{B}}(c_1)$ and using the same derivation as above. This ends the proof. $\square$
\end{proof}

To the best of the authors' knowledge, these theoretical results are new. It is important to note that the derivations have been conducted exclusively for the asymmetric setup, as extending them to the symmetric case appears to be prohibitively complex. These analytical expressions are expected to facilitate the optimization and comparison of practical schemes across a wide range of source and code parameters. 
For instance, optimizing $E_{\gamma}^{(q)}$, $E_{\nu}^{(qb)}$, and $A_{t}^{(qb)}$, could lead to an optimal quantize-binning scheme for DHT. We leave such optimization as future work.

\section{Numerical Results}\label{sec:numerical_results}
We now evaluate and compare the decision performance of the four proposed coding schemes: separate coding, truncation, quantization, and quantize-binning.
To this end, we provide Receiver Operating Characteristic (ROC)
curves which plot the Type-II error probability versus the Type-I error probability for each scheme.
ROC curves are a standard tool for evaluating the performance of hypothesis tests and provide a clear visualization of the trade-offs between Type-I and Type-II errors.


\subsection{Separate scheme versus truncation scheme}
We begin by comparing the two uncoded schemes:  the separate scheme and the truncation scheme. 
The source parameters are set to $c_0=0.1$, $p_1=0.5$ and $c_1=0.35$, while several values of $p_0 \in \{ 0.08, 0.2, 0.3\}$ are considered.  
We fix the source sequence length to $n=30$ bits, and the truncation length to $l=15$ bits.

\begin{figure}[t!]
\centerline{\includegraphics[width=0.48\textwidth]{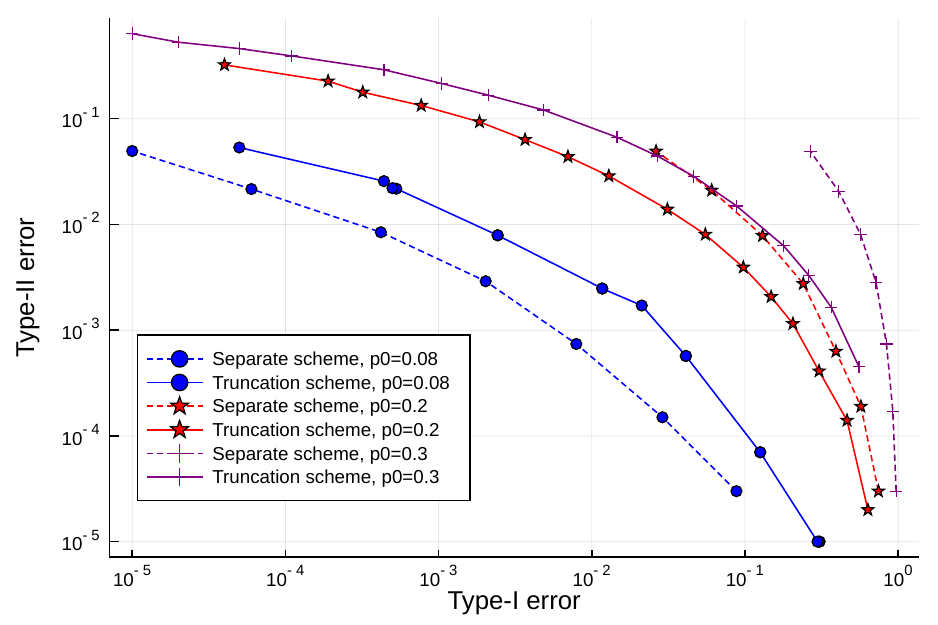}}
\caption{ROC curves for separate and truncation schemes.}
\label{fig:sep_vs_joint.eps}
\end{figure}

Figure~\ref{fig:sep_vs_joint.eps} shows the ROC curves of the two uncoded schemes for different values of $p_0$. It can be observed that when $p_0$ is small, the separate scheme outperforms the truncation scheme. This is because Encoder $1$ has access to the full $n$-bits sequence to decide between the two hypotheses characterized by significantly different values, $p_0 = 0.08$ and $p_1=0.5$. 
However, as $p_0$ increases and approaches $p_1$, the truncation scheme becomes advantageous. Indeed, in this regime, it becomes increasingly difficult for Encoder $1$ in the separate scheme to accurately distinguish between $\mathcal{H}_0$ and $\mathcal{H}_1$ based solely on its local observations. 
In contrast, the truncation scheme benefits from additional information through the values of $c_0$ and $c_1$,  which improves the decision-making process. 

In summary, while the separate scheme is rate-efficient, transmitting only one bit of information, its decision accuracy deteriorates as $p_0$ approaches $p_1$. Conversely, the truncation scheme, though less rate-efficient due to the transmission of truncated sequences at rates $R_1 = R_2 = R =l/n$, achieves better decision accuracy via joint decoding as $p_0$ increases. These observations motivate the development of coding schemes aimed at further improving both Type-I and Type-II error probabilities beyond the performance of the truncation scheme.

\subsection{Quantization scheme in the asymmetric setup} 
We now investigate the performance of the quantization
scheme compared to the truncation scheme in the asymmetric setup. 
We first assess the accuracy of the theoretical analysis  provided in Proposition~\ref{prop:error_quantization}. The source parameters are set to $p_0 = 0.5$, $p_1 = 0.5$, $c_1 = 0.5$, and $c_0 \in \{0.07,0.1\}$. We consider a source length $n=31$, and employ the ($31$,$16$) BCH code with minimum distance $d_{\min} = 7$. Consequently, a quantized vector of length $k=16$ is transmitted to the decoder. For a fair comparison, the truncation scheme is also set to transmit $l=16$ bits to the decoder.   

\begin{figure}[t!]
\centerline{\includegraphics[width=0.45\textwidth]{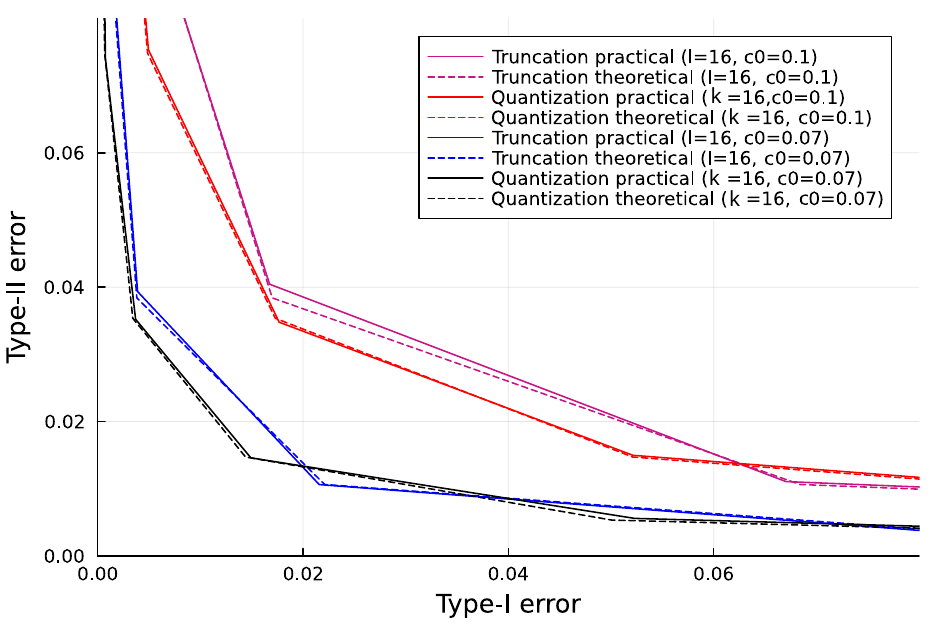}}
\caption{ROC curve for the $(31,16)$ BCH code used as a quantizer, compared to the truncation scheme in the asymmetric setup. }
\label{fig:tronc_vs_quant}
\end{figure}

Figure~\ref{fig:tronc_vs_quant} presents the ROC curves for both schemes with the previous parameters.
Theoretical curves are obtained via numerical evaluation of the expressions in Proposition~\ref{prop:error_quantization}, while practical results are measured from Monte Carlo simulations averaged over $10,000$ trials per point. 
The results indicate a close match between theoretical and empirical Type-I and Type-II error probabilities. 
 This is because the error probability expressions take into account the considered BCH code through the terms $E_\gamma^{(q)}$. Therefore, the theoretical expressions are found useful for the design of DHT curves. 

 Moreover, Figure~\ref{fig:tronc_vs_quant} 
  shows that the quantization scheme outperforms the truncation scheme in the asymmetric setup for the considered values of $c_0$, demonstrating the benefits of the proposed coding approach.


\begin{figure}[t!]
\centerline{\includegraphics[width=0.48\textwidth]{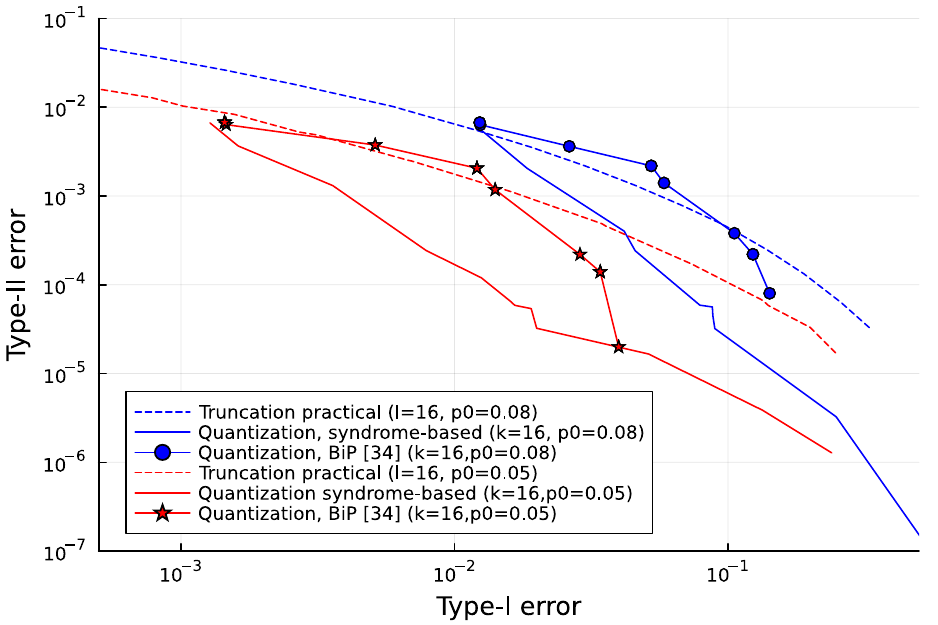}}
\caption{ROC curve for the $(31,16)$ BCH code used as a quantizer, compared to the truncation scheme in the symmetric setup. For the practical quantizer, we consider both our syndrome-based decoder and the BiP decoder from~\cite{castanheira2010lossy}}
\label{fig:3116_sym}
\end{figure}

\subsection{Quantization scheme in the symmetric setup} 
We now investigate the symmetric setup, aiming to compare the proposed quantization scheme against both the truncation scheme and existing quantization schemes based on LDGM codes with BiP algorithms~\cite{fridrich2007binary, wainwright2009low, castanheira2010lossy}. 
We first consider a source length $n=31$, with the $(31,16)$ BCH code. The source parameters are set to $p_0 \in \{ 0.05,0.08 \}$, $p_1 = 0.5$, $c_0 = 0.1$, $c_1 = 0.35$.  

Figure~\ref{fig:3116_sym} shows the ROC curves for the truncation scheme with $l=16$, the proposed quantization scheme with a syndrome-based decoder, and the quantization scheme using the BiP decoder with decimation as proposed in~\cite{castanheira2010lossy}.
A clear advantage is observed for the quantization scheme with the optimal syndrome-based decoder over the truncation scheme. 
The performance of the quantization scheme with the BiP decoder lies between that of the truncation scheme and the syndrome-based decoder. This performance gap is attributed to the fact that the BCH code is not sparse, which lowers the efficiency of message-passing algorithms such as BiP. Although LDGM codes could be considered to address this issue, their typically low minimum distance at short block lengths can severely degrade performance.
Finally, we observe that the curves for the quantization scheme in Figure~\ref{fig:3116_sym} have some points of irregularity. This comes from the test in equation~\eqref{eq:NP_criteria_quant_sym} which involves integer values  $w(\mathbf{x}_q^n)$ and $w(\mathbf{v}_q^n)$, and from the low source length $n=31$, which limits the range of these values.

\begin{figure}[t!]
\centerline{\includegraphics[width=0.5\textwidth]{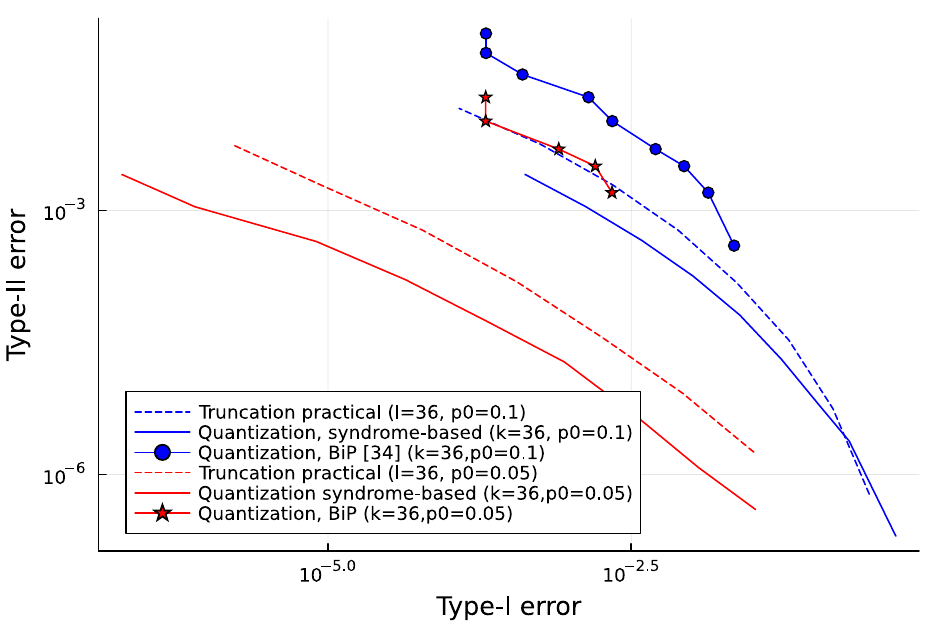}}
\caption{ROC curve for the $(63,36)$ BCH code used as a quantizer, compared to the truncation scheme in the symmetric setup. For the practical quantizer, we consider both the syndrome-based decoder and the BiP decoder from~\cite{castanheira2010lossy}. 
}
\label{fig:6336_sym}
\end{figure}

We next investigate the impact of code length in the symmetric setup. We consider a longer source sequence of $n=63$ bits and consider the $(63,36)$ BCH code with minimum distance $11$. The source parameters are set to $p_0 \in \{ 0.05,0.1 \}$, $p_1 = 0.5$, $c_0 = 0.1$, $c_1 = 0.35$. The ROC curves are shown in Figure~\ref{fig:6336_sym} for the truncation scheme, the quantization scheme using the syndrome-based decoder, and the quantization scheme using the BiP decoder with decimation of~\cite{castanheira2010lossy}. 
As before, a clear performance gain is observed for the quantization scheme with syndrome-based decoding over the truncation scheme. Furthermore, with the $(63,36)$ BCH code,  the quantization scheme using the BiP decoder exhibits a significant performance degradation compared to the other two schemes. 
We attribute this to the fact that the larger minimum distance of the $(63,36)$ BCH code makes the decoding problem increasingly more difficult to solve with the BiP algorithm.

\subsection{Quantize-binning scheme in the asymmetric setup}
We now evaluate the performance of the quantize-binning
scheme in the asymmetric setup. 
As with the quantization scheme, we first investigate the accuracy of the theoretical analysis of Proposition~\ref{prop:qandb}. The source parameters are set to $p_0=0.5$, $p_1 = 0.5$, $c_0 \in \{0.01, 0.03\}$, $c_1 = 0.35$. 
We consider a source length $n=31$, using the BCH $(31,16)$-code for quantization and the $(16,5)$ Reed-Muller code for binning. As a result, the final codeword length is  $\ell=8$ bits for the quantize-binning scheme, and $l=8$ bits for the truncation scheme. 

\begin{figure}[t!]
\centerline{\includegraphics[width=0.45\textwidth]{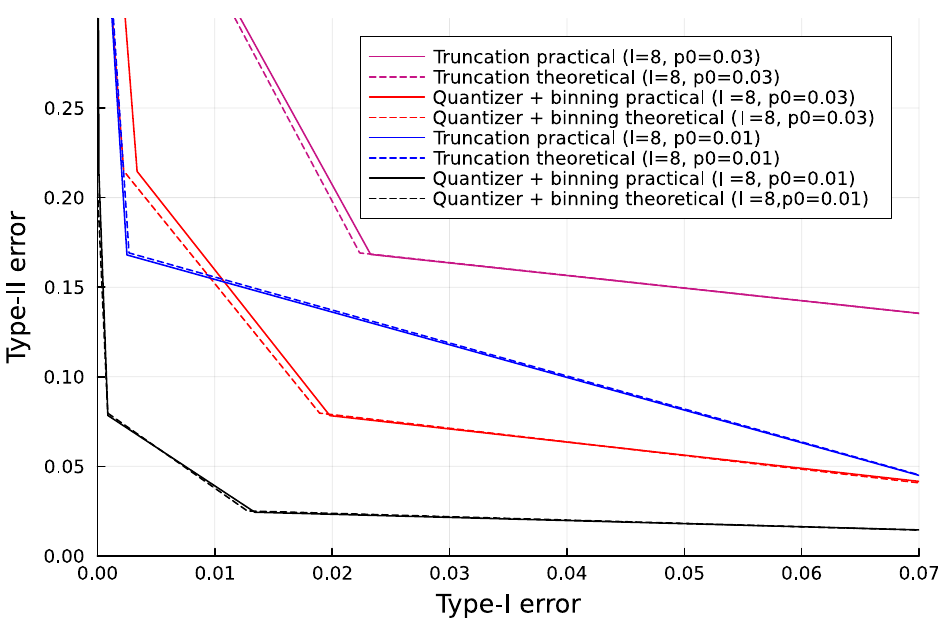}}
\caption{ROC curve for the quantize-binning scheme built from the $(31,16)$ BCH code  for quantization combined with the $(16,5)$ Reed-Muller code for binning in the asymmetric case.}
\label{fig:tronc_vs_quant_bin_asym}
\end{figure}

Figure~\ref{fig:tronc_vs_quant_bin_asym}
shows the ROC curves for both schemes. Theoretical curves are obtained by numerical evaluation of the expressions in  Proposition~\ref{prop:qandb}, while practical performance is measured via Monte Carlo simulations. 
The results show that the quantize-binning scheme provides a significant performance improvement over the truncation scheme. Furthermore, the theoretical Type-I and Type-II error probabilities closely match the empirical results, validating the accuracy of the theoretical analysis.
In summary, the quantize-binning scheme effectively improves the decision performance in the asymmetric setup while maintaining a low transmission rate. 

\subsection{Quantize-binning scheme in the symmetric setup}
\begin{figure}[t!]
\centerline{\includegraphics[width=0.5\textwidth]{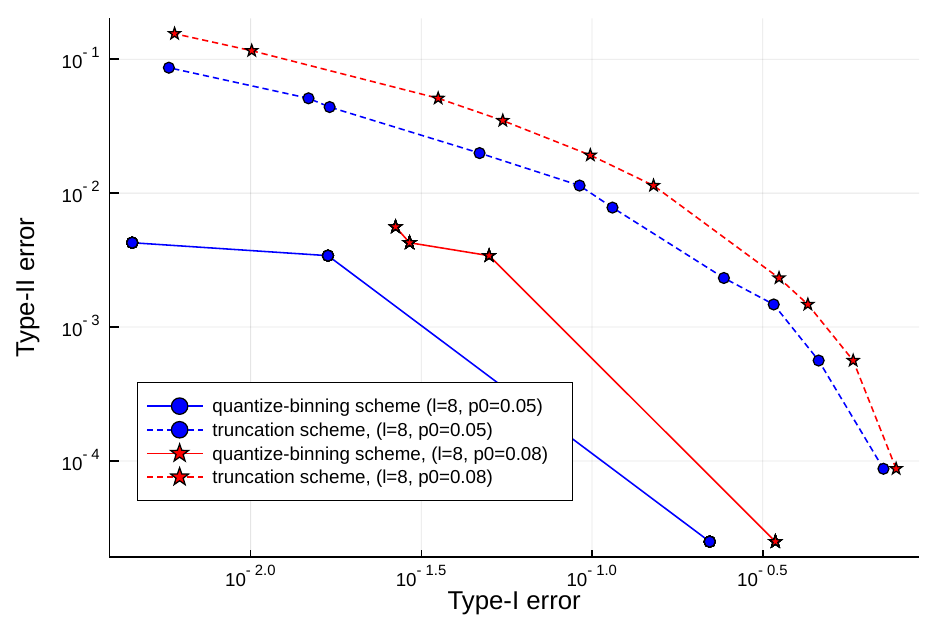}}
\caption{ROC curve for the quantize-binning scheme built from the BCH code $(31,16)$ for quantization combined with the Reed-Muller code $(16,5)$ for binning in the symmetric case.}
\label{fig:tronc_vs_quant_bin_sym}
\end{figure}

In the symmetric setup, the source parameters are set to $p_0=0.5$, $p_1 = 0.5$, $c_0 \in \{0.01, 0.03\}$, $c_1 = 0.35$.  We consider again the $(31,16)$ BCH code for the quantizers and the $(16,5)$-Reed-Muller code for the binning. Figure~\ref{fig:tronc_vs_quant_bin_sym} shows the ROC curves obtained from Monte-Carlo simulations, averaged over $10000$ trials for each point, for both the truncation scheme and the quantize-binning scheme. As before, we observe that the quantize-binning scheme outperforms the truncation scheme. These results confirm the relevance of practical quantization-binning schemes for DHT. 



\section{Conclusion}\label{sec:conc}
In this paper, we proposed practical short-length coding schemes for binary DHT in both asymmetric and symmetric setups.
We introduced two coding schemes, one built with a binary quantizer, and the other built as a quantize-binning scheme. Both schemes were constructed using short linear block codes. For each considered scheme, in addition to practical constructions, we derived theoretical expressions of Type-I and Type-II error probabilities in the asymmetric case.   Simulation results demonstrated that the proposed quantization and quantize-binning schemes outperform the baseline truncation scheme, and further confirmed the accuracy of the theoretical analyses. 
Future work will focus on leveraging the theoretical error probability expressions to optimize code parameters, and on considering more complex learning problems. 

%


\bibliographystyle{IEEEtran}
\bibliography{papers_done,sample}
\end{document}